\documentclass[11pt]{article}
\usepackage{latexsym}
\usepackage{amsmath} 

\begin{document}

\title{{\bf From Electromagnetic Duality \\
            to Extended Electrodynamics}}
\author{{\bf Stoil Donev}\\ Institute for Nuclear Research and Nuclear
Energy,\\ Bulg.Acad.Sci., 1784 Sofia, blvd.Tzarigradsko chausee 72\\
Bulgaria\\ e-mail: sdonev@inrne.bas.bg,\\}

\date{}

\maketitle

\begin{abstract}

This paper presents the transition from Classical Electrodynamics (CED) to
Extended Electrodynamics (EED) from the electromagnetic duality point of
view, and emphasizes the role of the canonical complex structure in ${\cal
R}^2$ in, both, nonrelativistic and relativistic formulations of CED and EED.
We begin with summarizing the motivations for passing to EED, as well as we
motivate and outline the way to be followed in pursuing the right extension
of Maxwell equations.  Further we give the nonrelativistic and relativistic
approaches to the extension and give explicitly the new equations as well as
some properties of the nonlinear vacuum solutions.

\end{abstract}

\section{Introduction}
Classical Electrodynamics is, obviously, not a linear theory in presence of
charges and currents. Indeed, the dynamics of the charge-carriers, considered
as a continuous system, is described by the following system of nonlinear
partial differential equations with respect to their velocity vector field
${\bf v}$,
$$
\mu\nabla_{\bf v} {\bf v}=\rho{\bf E}+\frac1c({\bf j}\times {\bf B}),
$$
or $u$ in the relativistic
formulation
$$
\mu c^2 u^\nu\nabla_{\nu} u_\sigma=-F_{\sigma\nu}j^\nu,
$$
where ${\bf j}=\rho {\bf v}$; $j=\rho u$, and $\mu$ is the invariant mass
density.  Since this velocity vector field
participates in the current expressions staying on the right-hand sides of
Maxwell equations:
\begin{equation}
\frac 1c \frac{\partial {\bf E}}{\partial t}=
{\rm rot}{\bf B} - \frac {4\pi}{c}
{\bf j}, \quad {\rm div}{\bf B}=0,                           
\end{equation}
\begin{equation}
\frac 1c \frac{\partial {\bf B}}{\partial t}=-{\rm rot}{\bf E},\quad \
\ {\rm div}{\bf E}=4\pi\rho,                                     
\end{equation}
or in relativistic notations
\begin{equation}
{\bf d}F=0,\quad {\bf d}*F=4\pi*j,                       
\end{equation}
then, obviously, the whole system becomes nonlinear. And we must consider the
whole system, otherwise the energy-momentum consrvation law will be violated.
Further, even in the pure field case, where no charges and currents are
present and the field equations are linear, Maxwell theory has its {\it
nonlinear part}, namely, the Poynting energy-momentum balance equation
\[
\frac1c\frac
{\partial}{\partial t} \frac {{\bf E}^2+{\bf B}^2}{8\pi}= -{\rm
div}\frac{{\bf E}\times{\bf B}}{4\pi},
\]
and this {\it nonlinear} equation is of
{\it basic importance} for the theory, even just because of its permanent and
everyday use. In fact, we should hardly trust Maxwell equations if this
everyday used and verified (for finite volume computations) Poynting equation
was not consistent with them. But, we should not forget that {\it the
Poynting balance equation may be consistent also with other field equations,
not just with Maxwell's ones, in particular, with appropriate nonlinear
ones}.  It is worth noting at this moment that the duality invariance as
considered in [1], is closely connected namely with the energy-momentum
quantities and relations, while Maxwell field equations may be cast into
$\mathcal{R}^2$ covariant form.

Let's recall now that the first steps of theoretical physics, made by Newton
with his three laws of mechanics, establish, in fact, the fundamental
conservation laws in the form of differential balance equations for the
momentum and energy.

The {\it universal} and {\it conservative} character of these two quantities,
energy and momentum, explains the remarkable power of classical mechanics. As
for their importance in quantum theory we could hardly imagine it without
{\it the Hamiltonian}.  The conservation laws are the heart of all physics,
because physics is doing with real objects, and we could hardly think of real
physical objects at all if these objects have no any constant in time
properties.  Further, we could hardly understand the interaction in nature if
there are no {\it universal} (i.e. carried by any physical object)
conservative quantities like {\it energy} and {\it momentum}.  Indeed,
from modern physics point of view interaction in mechanics, as well as in the
other branches of physics, {\it necessarily} requires energy-momentum
exchange. The differential equation form of this energy-momentum exchange
gathers together the two mutually consistent tendencies of existence:  {\it
conservation} and {\it alteration}.  And the {\it alteration}, or {\it
time-evolution}, defines, in fact, those boundaries behind which the physical
system under consideration can not exist anymore.  This duality between
conservation and time-evolution is theoretically implemented through the
{\it dynamical equations} of the physical system under consideration, and
only those dynamical equations have to be considered as {\it reasonable}
whose solutions have {\it reasonable} conservation and stability properties,
e.g.  the evolution of a free and time-stable object must not lead to a
self-ruin, and the corresponding conservative quantities carried by the
solutions should accept {\it finite} values, {\it not} infinite ones.


In view of the above we can look at the second principle of Newton in the
following way:

{\bf The basic equations, governing some class of
mechanical objects and their interactions, should start with establishing how
the local energy-momentum exchange among these objects is performed, and all
further peculiarities of their behavior to appear in the theory as
correspondingly consistent relations with this basic initial fundament}.

Looked at this way this Newton's principle can be easily and
appropriately extended to description of continuous (field) objects.  We
know that the local energy-momentum conservation laws of every (linear or
nonlinear) field theory are {\it nonlinear} partial differential equations,
and, following the Newton's approach of pointing out dynamical equations, we
must pay the corresponding respect these nonlinear equations deserve.  In
other words, we should establish first {\it how much} and {\it in what way}
the continuous physical system under consideration is potentially able to
exchange locally energy-momentum with the rest of the world, and afterwards
to go on with taking into account in a consistent way its other features.
This is the general approach we are going to follow in looking for an
adequate nonlinearization of the pure field Maxwell equations.

On the other hand, it is much easier to work with linear dynamical field
equations, especially if we have some corresponding experimental evidence for
the assumption of such linear equations as a theoretical basis. But we must
{\it not forget} the {\it limited} character of {\it any} specific
experimental evidence when it is considered as a basis for fundamental
assumptions. It seems more reliable to establish first the local relations,
describing the balance of at least some of the universal conserved
quantities, and then to go further with more precisions and specifications.

Let's recall now some features of the classical electromagnetic pure field
theory.  As it is well known, it is traditionally taught starting with
Faraday's electromagnetic induction law and with Maxwell's magnetoelectric
induction law :
\[
\frac {d}{dt}\int_S* {\bf B}=-c\int_l{\bf E}, \quad \frac
{d}{dt}\int_S*{\bf E}=c\int_l{\bf B}.
\]
\vskip 0.5cm
\noindent
{\bf Remark}: Here and further we denote by $*$ the Hodge operator, defined by
the corresponding metric $g$ and volume form $\omega_g$, through the
relation
\[
\alpha\wedge\beta=g(*\alpha,\beta)\omega_g,
\]
where $\alpha$ and $\beta$ are $p$ and $n-p$ forms respectively.
Also, we identify through the metric the covariant and contravariant
vector and tensor fields.
\vskip 0.5cm
\noindent
The above integral relations, together with
${\rm div}{\bf E}=0$ and ${\rm div}{\bf B}=0$ lead to linear field equations
for the components of ${\bf E}$ and ${\bf B}$, moreover, as a necessary
condition it is obtained that any component $U$ of these vector fields is
obliged to satisfy the d'Alembert wave equation $\Box U=0$. Now, the
solutions of the field equations are meant to describe real time-stable
continuous objects, usually called fields.  Unfortunately, the wave equation
$\Box U=0$ predicts {\it strong time-instability} for any smooth enough {\it
finite} initial condition (the Poisson's theorem), and besides, the {\it
infinite} initial conditions (e.g. those leading to harmonic plane waves)
require {\it infinite} energy of the corresponding solutions. It seems hardly
reasonable to think of real objects carrying infinite energy, so we have to
admit that the free field solutions of the Faraday-Maxwell equations do not
present adequate enough models of any real continuous object because of the
{\it finite} and {\it time-stable} nature of the latter.

In order to see the merits of the Poynting balance equation in this relation
we consider first the well known plane wave solution of the pure field Maxwell
equations in the appropriate coordinate system:
\[
{\bf E}=\Bigl[u(\xi+\varepsilon z), p(\xi+\varepsilon z), 0\Bigr],\
{\bf B}=\Bigl[\varepsilon p(\xi+\varepsilon z),
-\varepsilon u(\xi+\varepsilon z), 0\Bigr], \varepsilon=\pm 1, \xi=ct,
\]
where $u$ and $p$ are arbitrary differentiable functions. Even if $u$ and $p$
are soliton-like with respect to the coordinate $z$, they do not depend on
the other two spatial coordinates $(x,y)$. Hence, the solution occupies the
whole ${\cal R}^3$, or its infinite subregion, and clearly it carries
infinite integral energy
\[
W=\frac{1}{4\pi}\int_{{\cal R}^3}\frac{{\bf E}^2+{\bf B}^2}{2}dxdydz=
\frac{1}{4\pi}\int_{{\cal R}^3}(u^2+p^2)dxdydz=\infty.
\]
In particular, the popular harmonic plane wave
$$
u=U_o\cos (\omega t\pm k_z.z),\ p=P_o \sin (\omega t\pm k_z.z),\
c^2 k_z^2=\omega^2,\ U_o=const,\ P_o=const,
$$
clearly occupies the whole 3-space and carries infinite energy
$$
W=\frac{1}{4\pi}\int_{{\cal R}^3}(U_o+P_o)dxdydz =\infty.
$$
The plane wave solutions reflect well enough some features of the notion for
energy-momentum propagation in a fixed spatial direction (the axis $z$ in
this system of coordinates), but they all are infinite, no dependence on the
transverse coordinates ($(x,y)$ in this system of coordinates) is allowed by
the equations, just dependence on the running wave argument $(\xi+\varepsilon
z)$ of $u$ and $p$ is allowed.

Following the Newton's approach, let's check now the pure field Poynting
equation, which is nonlinear and which describes {\it differentially} the
local intrafield energy-momentum redistribution during the time evolution,
whether it admits finite, i.e. spatial soliton-like solutions. Suppose
that in the above system of coordinates we have $u=u(x,y,\xi+\varepsilon z)$
and $p=p(x,y, \xi+\varepsilon z)$, where the dependence on the three spatial
coordinates is {\it arbitrary}. The corresponding ${\bf E}$ and ${\bf B}$
surely do {\it not} define solution to Maxwell equations, and we check if
these ${\bf E}$ and ${\bf B}$ define solution to the Poynting equation.
We obtain $\frac12({\bf E}^2+{\bf B}^2)=u^2+p^2$ and
${\bf E}\times{\bf B}=-\varepsilon(0,0,u^2+p^2)$. Denoting
$u^2+p^2=\phi$ and the derivative of $\phi$ with respect to $(\xi+\varepsilon
z)$ by $\prime$ we obtain
\[
\frac{\partial }{\partial t}\Phi=c\Phi',\
-c{\rm div}({\bf E}\times{\bf B})=c\Phi',
\]
i.e. the Poynting equation is satisfied. Now, since $u$ and $p$ are
{\it arbitrary}
functions of their arguments we conclude that the Poynting equation {\it
does} admit photon-like (3+1) soliton solutions, while Maxwell equations
do NOT, they predict a quick self-ruin of any finite 3-dimensional smooth
enough initial field configuration. May be it is worth at this moment to say
that under (3+1)-soliton we meen a {\it time stable continuous nondispersing
finite object of 3 spatial dimensions, having internal dynamical structure,
carrying finite integral energy-momentum, and the translational component of
its propagation velocity is constant and is along some (straightline)
direction in the 3-space}.  Such solitons are called {\it photon-like} if
they move translationally as a whole with the velocity of light, which meens
that their integral energy-momentum vector has zero length in Minkowski
space-time. The corresponding solutions, describing such objects, I call
{\it (3+1) soliton-like} or just {\it soliton} solutions.
\vskip 0.5cm

Hence, we are facing two alternatives: the {\it first} one offers {\it
linear} equations with nonreasonable, i.e infinite, or finite but strongly
time-unstable, vacuum solutions; the {\it second} one {\it could} offer
reasonable finite and time-stable vacuum solutions if an {\it appropriate
nonlinearization} of the linear Maxwell equations, consistent with the
Poynting relation, is found.

Another reason to reconsider from this point of view classical
Faraday-Maxwell theory comes from quantum theory, where the elementary
quantum objects (free photons, free electrons, etc.) are considered as
point-like, i.e. {\it structureless}, objects as it is in classical
mechanics.  But the Planck's formula $E=h\nu$ definitely {\it requires} any
of these free objects to demonstrate {\it intrinsic periodic process} with
frequency $\nu$, which {\it no point-like free object can do}: if the object
has no structure the periodicity may be caused only by an outside agent,
which meens that the object is NOT free.  Further, this {\it no structure
assumption} makes the elementary quantum objects {\it eternal} and {\it
undestroyable}, because there is nothing, no structure, to be destroyed; they
are allowed just to change their energy-momentum under any external
perturbation, therefore, no explanation of the observed transformations
(e.g.  anihilation) among these microobjects under collisions would be
possible:  {\it undestroyable entities can not transform into each other}. In
short, the free elementary quantum objects do {\it not admit in principle}
the point-like, i.e. the structureless, approximation, and this seems to be
the {\it most important difference} between {\it classical} objects and {\it
quantum} objects.  Therefore, it seems unreasonable to try to build theory of
quantum objects on the assumption that they are considered as classical (i.e.
nonquantum) objects.

In our view, we have to let such inconsistencies go out of physical theories,
and we must pay the corresponding respect to the structure these microobjects
possess through some further reasonable and appropriate development of the
theory. Therefore, any success of modern theoretical physics in doing with
extended, {\it not} point-like, field objects, must be correspondingly
respected and appreciated. In view of no enough initial experience and
insight in working with such objects, it seems more reasonable to begin with
a study of a specific field, than starting a straightforward attack of the
most general case. The comparetively well developed classical and quantum
electrodynamics make photons the most natural and the most promising objects
for this purpose.

Considerations of this kind made us favor the second of the above
mentioned two alternatives. In what follows we shall briefly outline our
approach to nonlinearization of Maxwell equations which was called Extended
Electrodyanmics [2].
The suggestive nature of the duality properties of the electromagnetic field,
which were appropriately interpreted and described in [1], will be clearly
seen and thoroughly used.

\vskip 0.5cm

\section {Nonrelativistic consideration}
As it was shown in [1] the dual nature of the electromagnetic field naturally
leads to the nonrelativistic formulation of Maxwell equations by means of the
${\cal R}^2$ valued 1-form $\omega$ and of the canonical complex structure
${\cal I}$ of ${\cal R}^2$ (we use the notations in [1]):

\begin{equation}
\omega={\bf E}\otimes\varepsilon^1 + {\bf B}\otimes\varepsilon^2,   
\end{equation}
\begin{equation}
*{\bf d}\omega-\frac {1}{c} \frac {\partial }{\partial t}{\cal I}_*
(\omega)=\frac {4\pi}{c}{\cal I}_*({\cal J}),                       
\ \ \delta \omega =-4\pi {\cal Q},
\end{equation}
\begin{equation}
{\cal Q}=\rho_e\otimes \varepsilon^1+\rho_m\otimes \varepsilon^2,\ \
{\cal J}={\bf j}_e\otimes \varepsilon^1+{\bf j}_m\otimes \varepsilon^2,
\end{equation}
where $\delta$ is the coderivative, $(\varepsilon^1,\varepsilon^2)$ is the
canonical basis of ${\cal R}^2$,
$$
{\cal I}_*\omega=
{\bf E}\otimes{\cal I}(\varepsilon^1)+
{\bf B}\otimes{\cal I}(\varepsilon^2)=
{\bf E}\otimes\varepsilon^2-{\bf B}\otimes\varepsilon^1,
$$
and $\rho_e, \rho_m, {\bf j}_e, {\bf j}_m$ are  electric density, magnetic
density, electric current and magnetic current, respectively.

The above formulae imply that the field has two differentially
interrelated but algebraically distinguished components. Following our
approach, we look now on this circumstance from energy-momentum exchange
point of view.  Clearly, the field $\omega$ is potentially able to exchange
energy-momentum with any other physical system, which also is potentially
able to exchange energy-momentum with the field, and this
energy-momentum exchange may, in general, be carried out through each of its
2 vector components ${\bf E}$ and ${\bf B}$.  Further, an intra-field
energy-momentum exchange between the two vector components of the field may
also take place, and this third intra-field exchange may be responsible for
the intrinsic spin momentum of the field.  Hence, we have to describe these
{\it three} potentially possible and independent exchange processes, where
{\it independent} means that, in general, any of these exchanges may go
without the other two to occur. In particular, the intra-field exchange may
occur in the pure field case.

Such a situation can be modelled by pointing out a 3-dimensional vector space
$W$, where each of the dimensions will account for only one of these exchange
processes.  Since our field object $\omega$ takes values in ${\cal R}^2$ it
is natural to try to connect algebraically $W$ with ${\cal R}^2$.  The
simplest such space appears to be the symmetrized tensor product
$Sym({\cal R}^2\otimes{\cal R}^2)={\cal R}^2\vee{\cal R}^2$,
which is 3-dimensional. So, since
$\vee:{\cal R}^2\times {\cal R}^2\rightarrow {\cal R}^2\vee{\cal R}^2$
is a bilinear map we can multiply two ${\cal R}^2$ valued differential forms
through $\vee$, as it was pointed out in [1]. The three components of the
obtained ${\cal R}^2\vee{\cal R}^2$ valued differential form are meant to
represent the densities of the above mentioned energy-momentum exchange
quantities.

We begin working out this idea through equations (5), of course.
Let's multiply the left-hand side of the first (5) equation from the
right by $\omega$ through $\vee$ and take the euclidean $*$ from the left. We
obtain
\begin{equation*}
	\begin{split}
&
*\vee\left(*{\bf d}\omega -\frac1c\frac{\partial }{\partial t}{\cal
I}_*\omega,\omega\right)=
\Biggl[\left({\rm rot}{\bf E}+\frac1c\frac{\partial
{\bf B}}{\partial t}\right)\times
{\bf E}\Biggr]\otimes\varepsilon^1\vee\varepsilon^1
\\
&
+\Biggl[\left({\rm rot}{\bf B}-\frac1c\frac{\partial
{\bf E}}{\partial t}\right)\times{\bf
B}\Biggr]                                                     
\otimes\varepsilon^2\vee\varepsilon^2
+\Biggl[\left({\rm rot}{\bf B}-\frac1c\frac{\partial
{\bf E}}{\partial t}\right)\times{\bf E}+
\left({\rm rot}{\bf E}+\frac1c\frac{\partial
{\bf B}}{\partial t}\right)\times{\bf
B}\Biggr]\otimes\varepsilon^1\vee\varepsilon^2
	\end{split}
\end{equation*}
Now we take ${\cal I}_*$ from the left of the second (5) equation and
multiply from the right by $-{\cal I}_*\omega$ through $\vee$. We obtain

\[
-\vee\left({\cal
I}_*\delta \omega,{\cal I}_*\omega\right)= {\bf B}{\rm div}{\bf
B}\otimes\varepsilon^1\vee\varepsilon^1+ {\bf E}{\rm div}{\bf
E}\otimes\varepsilon^2\vee\varepsilon^2- \left({\bf B}{\rm div}{\bf E}+{\bf
E}{\rm div}{\bf B}\right)\otimes \varepsilon^1\vee\varepsilon^2.
\]
We sum up now these two relations:

\[ *\vee\left(*{\bf d}\omega
-\frac1c\frac{\partial }{\partial t}{\cal I}_*\omega,\omega\right)-
\vee\left({\cal I}_*\delta \omega,{\cal I}_*\omega\right)=
\]
\[
\Biggl[\left({\rm rot}{\bf E}+\frac1c\frac{\partial
{\bf B}}{\partial t}\right)\times
{\bf E}+{\bf B}{\rm div}{\bf B}\Biggr]\otimes\varepsilon^1\vee\varepsilon^1+
\Biggl[\left({\rm rot}{\bf B}-\frac1c\frac{\partial
{\bf E}}{\partial t}\right)\times{\bf
B}+{\bf E}{\rm div}{\bf E}\Biggr]\otimes\varepsilon^2\vee\varepsilon^2+
\]
\[
+\Biggl[\left({\rm rot}{\bf B}-\frac1c\frac{\partial
{\bf E}}{\partial t}\right)\times{\bf E}+
\left({\rm rot}{\bf E}+\frac1c\frac{\partial
{\bf B}}{\partial t}\right)\times{\bf
B}-{\bf B}{\rm div}{\bf E}-{\bf E}{\rm div}{\bf B}
\Biggr]\otimes\varepsilon^1\vee\varepsilon^2.
\]
Now we do the same operations with the right-hand sides of equations (5) and
obtain
\[
\frac{4\pi}{c}*\vee \left({\cal I}_*{\cal J},\omega\right)+
4\pi\vee\left({\cal I}_*{\cal Q},{\cal I}_*\omega\right)=
\]
\[
=4\pi\Biggl[\left(\frac1c\left(-{\bf j}_m\times{\bf E}\right)
+\rho_m{\bf B}\right)\otimes
\varepsilon^1\vee\varepsilon^1+
\left(\frac1c\left({\bf j}_e\times{\bf B}\right)+\rho_e{\bf E}\right)\otimes
\varepsilon^2\vee\varepsilon^2+
\]
\[
+\Bigl(\frac1c\left(-{\bf j}_m\times{\bf B}
+{\bf j}_e\times{\bf E}\right)-
\rho_m{\bf E}-\rho_e{\bf B}\Bigr)\otimes
\varepsilon^1\vee\varepsilon^2\Biggr]
\]
Hence, the nonlinear equation
\begin{equation}
*\vee\left(*{\bf d}\omega
-\frac1c\frac{\partial }{\partial t}{\cal I}_*\omega,\omega\right)-
\vee\left({\cal I}_*\delta \omega,{\cal I}_*\omega\right)=
\frac{4\pi}{c}*\vee \left({\cal I}_*{\cal J},\omega\right)+         
4\pi\vee\left({\cal I}_*{\cal Q},{\cal I}_*\omega\right)
\end{equation}
gives the following three equations
\begin{equation}
\left({\rm rot}{\bf E}+\frac1c\frac{\partial
{\bf B}}{\partial t}\right)\times                            
{\bf E}+{\bf B}{\rm div}{\bf B}=
\frac{4\pi}{c}\left(-{\bf j}_m\times{\bf E}\right)
+4\pi\rho_m{\bf B},
\end{equation}
\begin{equation}
\left({\rm rot}{\bf B}-\frac1c\frac{\partial
{\bf E}}{\partial t}\right)\times                             
{\bf B}+{\bf E}{\rm div}{\bf E}=
\frac{4\pi}{c}\left({\bf j}_e\times{\bf B}\right)+
4\pi\rho_e{\bf E},
\end{equation}

\begin{equation}
	\begin{split}
&\left({\rm rot}{\bf B}-\frac1c\frac{\partial
{\bf E}}{\partial t}\right)\times{\bf E}+
\left({\rm rot}{\bf E}+\frac1c\frac{\partial              
{\bf B}}{\partial t}\right)\times{\bf B}-
{\bf B}{\rm div}{\bf E}-{\bf E}{\rm div}{\bf B}=
\\
&=\frac{4\pi}{c}\left(-{\bf j}_m\times{\bf B}+{\bf j}_e\times{\bf E}\right)-
4\pi\rho_m{\bf E}-4\pi\rho_e{\bf B}.
	\end{split}
\end{equation}

It is clearly seen that the right-hand side of equation (10) becomes zero
every time when the electric and magnetic currents and charges are zero, so
it depends algebraically on the energy-momentum exchanges described by the
first two equations, and this is not in a full accordance with our assumption
that all of the three exchanges are independent on each other. Besides,
generally speaking, we could imagine that there exist some new, unknown kind
of media built not of electric and magnetic charges as in the classical case,
but still able to exchange energy momentum with the field, e.g. a
gravitational field. This suggests the following formal generalization:  4
algebraically independent vector fields ${\bf a}^i, i=1,2,3,4$ and 4
functions $a^i, i=1,2,3,4$  to be introduced, so that in the most general
case we shall have
\begin{equation}
\left({\rm rot}{\bf B}-\frac1c\frac{\partial
{\bf E}}{\partial t}\right)\times                            
{\bf B}+{\bf E}{\rm div}{\bf E}=
{\bf a}^1\times{\bf B}+a^1{\bf E}
\end{equation}
\begin{equation}
\left({\rm rot}{\bf E}+\frac1c\frac{\partial
{\bf B}}{\partial t}\right)\times                             
{\bf E}+{\bf B}{\rm div}{\bf B}=
{\bf a}^4\times{\bf E}-a^4{\bf B}
\end{equation}
\begin{equation}
\begin{split}
&\left({\rm rot}{\bf B}-\frac1c\frac{\partial
{\bf E}}{\partial t}\right)\times{\bf E}+
\left({\rm rot}{\bf E}+\frac1c\frac{\partial              
{\bf B}}{\partial t}\right)\times{\bf B}-
{\bf B}{\rm div}{\bf E}-{\bf E}{\rm div}{\bf B}=
\\
&={\bf a}^2\times{\bf B}+{\bf a}^3\times{\bf E}+
a^2{\bf E}-a^3{\bf B}.
\end{split}
\end{equation}
Obviously, equations (8)-(10) correspond to
\begin{equation}
{\bf a}^1={\bf a}^3=\frac{4\pi}{c}{\bf j}_e;\
{\bf a}^2={\bf a}^4=\frac{4\pi}{c}{\bf j}_m;\           
a^1=-a^3=4\pi\rho_e;\ a^2=a^4=-4\pi\rho_m.
\end{equation}
We have to say, that, in general, the vector fields ${\bf a}^i$ and the
corresponding functions $a^i$ are not subject to the condition to satisfy
corresponding continuity equations, although this is not forbidden, this
depends on the features of the medium which participates in the
energy-momentum exchange with the field through its own ${\bf a}^i, a^i$.

We mention that choosing appropriately the quantities $({\bf a}^i, a^i)$
in our equations we can obtain, for example, the extension of
Maxwell pure field equations considered by B.Lehnert [3]. In fact, if
we put ${\bf a}^2={\bf a}^4=0$, $a^2=a^4=0$,  ${\rm div}{\bf B}=0$,
${\bf a}^1={\bf a}^3=const.{\bf j}$, $a^1=a^3={\rm div}{\bf E}=\sigma$,
${\bf j}=\sigma{\bf C}$, and ${\bf C}^2=c^2$, where $c$ is the
velocity of light in vacuum, then the solutions of Lehnert's extension of
Maxwell equations constitute a class of solutions of our equations. The only
difference is in the interpretation:  B.Lehnert consideres these equations as
free-field, while in our approach they will appear as describing some special
kind of medium.

Let's write down explicitly the corresponding vacuum equations:
\begin{equation}
\left({\rm rot}{\bf B}-\frac1c\frac{\partial
{\bf E}}{\partial t}\right)\times                            
{\bf B}+{\bf E}{\rm div}{\bf E}=0
\end{equation}
\begin{equation}
\left({\rm rot}{\bf E}+\frac1c\frac{\partial
{\bf B}}{\partial t}\right)\times                             
{\bf E}+{\bf B}{\rm div}{\bf B}=0
\end{equation}
\begin{equation}
\left({\rm rot}{\bf B}-\frac1c\frac{\partial
{\bf E}}{\partial t}\right)\times{\bf E}+
\left({\rm rot}{\bf E}+\frac1c\frac{\partial              
{\bf B}}{\partial t}\right)\times{\bf B}-
{\bf B}{\rm div}{\bf E}-{\bf E}{\rm div}{\bf B}=0.
\end{equation}

We note, that Faraday-Maxwell theory separates its own sector of solutions in
the frame our more general nonlinear approach based on equations (11)-(13),
nothing from that theory is lost and may be used in every situation where it
is considered to provide a good enough approximation.  But, for example, in
case of describing soliton-like behavior of electromagnetic radiation in free
space we have to turn to the nonlinear sector of solutions of (15)-(17).
As for the consistency of the pure field equations (15)-(17) with the
Poynting pure field equation, it follows from (15)-(17) that
\begin{equation}
{\bf E}.\left({\rm rot}{\bf B}-
\frac1c\frac{\partial {\bf E}}{\partial t}\right)=0,\
{\bf B}.\left({\rm rot}{\bf E}+                              
\frac1c\frac{\partial {\bf B}}{\partial t}\right)=0,
\end{equation}
and from these two relations the Poynting pure field equation follows.

Finally, in order to get some initial impression about the nonlinear
vacuum solutions of (15)-(17) we note some of their properties. First,
talking about nonlinear vacuum solutions of (15)-(17) we mean those
vacuum solutions which satisfy the following nonequalities:
\begin{equation}
{\rm rot}{\bf E}+\frac{\partial {\bf B}}{\partial \xi}\neq 0, \ \ {\rm
rot}{\bf B}-\frac{\partial {\bf E}}{\partial \xi}\neq 0, \ \ {\bf div}{\bf E}
\neq 0,\ \ {\bf div}{\bf B}\neq 0.                 
\end{equation}
Now it is easy to verify that, besides (18), the nonlinear vacuum solutions
satisfy the following relations:
\[
{\bf E}.{\bf B}=0,\
{\bf E}^2={\bf B}^2,\ {\bf B}.\left({\rm rot}{\bf B}- \frac1c\frac{\partial
{\bf E}}{\partial t}\right)= {\bf E}.\left({\rm rot}{\bf E}+
\frac1c\frac{\partial {\bf B}}{\partial t}\right).
\]
These relations guarantee that $|{\bf E}|$ and $|{\bf B}|$ of the
nonlinear solutions of (15)-(17) are invariant with respect to the duality
transformations.

Consider now the vector fields
\begin{equation}
\vec{\cal E}={\rm rot}{\bf E}+\frac{\partial {\bf B}}{\partial \xi}+
\frac{{\bf E}\times
{\bf B}}{|{\bf E}\times {\bf B}|}{\rm div}{\bf B},   
\end{equation}
\begin{equation}
\vec{\cal B}=
{\rm rot}{\mathbf B}-\frac{\partial {\mathbf E}}{\partial \xi}-
\frac{{\bf E}\times
{\mathbf B}}{|{\mathbf E}\times {\mathbf B}|}{\rm div}{\mathbf E}.    
\end{equation}
Under a duality transformation they transform like ${\bf E}$ and ${\bf B}$
respectively. Equations (15)-(17) are equivalent respectively to
\begin{equation}
\vec{\cal E}\times{\bf E}=0,\ \vec{\cal B}\times{\bf B}=0,\          
\vec{\cal E}\times{\bf B}+\vec{\cal B}\times{\bf E}=0.
\end{equation}
It follows:
\[
\vec{\cal E}=f{\bf E},\ \vec{\cal B}=f{\bf B},\
|\vec{\cal E}|=|\vec{\cal B}|,
\]
where $f$ is a function. These properties allow the important concept of {\it
scale factor} $L({\bf E},{\bf B})$ for any nonlinear vacuum solution to be
defined by
\begin{equation}
L({\mathbf E},{\mathbf B})=\frac{1}{|f|}=
\frac{|{\mathbf E}|}{|\vec{\cal E}|}=                        
\frac{|{\mathbf B}|}{|\vec{\cal B}|}.
\end{equation}
Clearly, $L({\bf E},{\bf B})$ is also invariant with respect to duality
transformations. Hence, every nonlinear solution defines its own scale.

We examine now the vector fields $\vec{\cal E}$ and $\vec{\cal B}$ as
measures of the internal (spin) angular momentum. Clearly, we have in view
only the nonlinear solutions of (15)-(17).  The third equation of (23) says
that $\vec{\cal E}\times {\bf B}=- \vec{\cal B}\times {\bf E}$. Both sides of
this relation measure the same changes of momentum: the left-hand side says
how much momentum is transferred from ${\bf E}$ to ${\bf B}$, and the
right-hand side says how much momentum is transferred from ${\bf B}$ to ${\bf
E}$, and these quantities are equal in magnitude. Note that these mutual
transfers are made through the rotations ${\rm rot}{\bf E}$ and ${\rm
rot}{\bf B}$ of ${\bf E}$ and ${\bf B}$, so we could interpret these
continuous processes of intrafield momentum transfers as an appearance of the
intrinsic rotational (spin) features of the solution. Further, besides these
rotational degrees of freedom, the nature of the solution makes the field
propagate along the spatial direction defined by the Poynting vector, so it
carries energy-momentum along ${\bf E}\times{\bf B}$.  Therefore, we may
expect the time derivative of the projection of $\vec{\cal E}\times{\bf B}$
on the direction of propagation to be equal to ${\rm div}(\vec{\cal
E}\times{\bf B})$.  Hence, we may assume
\begin{equation}
\frac1c\frac{\partial }{\partial t}\frac{(\vec{\cal E}\times{\bf B}).
({\bf E}\times{\bf B})}{|{\bf E}\times{\bf B}|}=                   
-{\rm div}(\vec{\cal E}\times{\bf B}).
\end{equation}
In other words, we consider the quantity
\begin{equation}
\vec{\cal H}=(\vec{\cal E}\times{\bf B}).\frac
{{\bf E}\times{\bf B}}{|{\bf E}\times{\bf B}|}                  
\end{equation}
as a measure of the spin momentum density, so its time change has to come from
somewhere, and equation (24) defines this change as the divergence of
$\vec{\cal E}\times{\bf B}=-\vec{\cal B}\times{\bf E}$. Equation (24) has to
be considered, of course, as an additional relation to equations (15)-(17)
when these intra-rotational properties of the solutions are considered as
important enough to be quantitatively accounted for.

\vskip 0.5cm

\section{Relativistic consideration}

We recall from [1] the concept of $(*,{\cal I})$-equivariant ${\cal R}^2$
valued differential form $\Gamma$:
$$
(*,{\cal I})\Gamma=*\Gamma_1\otimes{\cal I}(\varepsilon^1) +
		 *\Gamma_2\otimes{\cal I}(\varepsilon^2)=
\Gamma_1\otimes\varepsilon^1+\Gamma_2\otimes\varepsilon^2=\Gamma,
$$
where $(\varepsilon^1,\varepsilon^2)$ is the canonical basis of ${\cal R}^2$.
Further we consider Minkowski space-time with metric tensor
$\eta_{\mu\nu}$, signature $(-,-,-,+)$ and corresponding $*$-operator.

The relativistic nonlinearization respects the following principles:

-the basic field is ${\cal R}^2$-valued $(*,{\cal I})$-equivariant
differential 2-form $\Omega$ on Minkowski space-time,

-the basic differential vacuum equations for $\Omega$ must appear as one
3-dimensional relation for one object,

-the physical sense of the basic differential equations in presence of
external fields must be local balance of universal conserved quantities,

-Faraday-Maxwell theory must be entirely present in the nonlinear
generalization.

The requirement for $(*,{\cal I})$-equivariance of $\Omega$, where
$\mathcal{I}$ is the canonical complex structure in $\mathcal{R}^2$,
implies that
in the canonical orthonormal basis $(\varepsilon^1,\varepsilon^2)$ in
${\cal R}^2$ we have
$\Omega=F\otimes\varepsilon^1+*F\otimes\varepsilon^2$.
 Now, these basis vectors determine two subspaces
$\{\varepsilon^1\}$ and $\{\varepsilon^2\}$ of $\mathcal{R}^2$ and the
corresponding projections
$pr_{1}:\mathcal{R}^2\rightarrow\{\varepsilon^1\};\
pr_{2}:\mathcal{R}^2\rightarrow\{\varepsilon^2\}$, $\mathcal{R}^2=
\{\varepsilon^1\}\oplus\{\varepsilon^2\}$. These two projection operators
extend to projections $\pi_1$ and $\pi_2$
in the $\mathcal{R}^2$-valued differential forms on $M$:
\[
\pi_1 \Omega =\pi_1(\Omega^1\otimes k_1+\Omega^2 \otimes k_2)=
\Omega^1 \otimes \pi_1 k_1 +\Omega^2 \otimes \pi_1 k_2=
\]
\[
=\Omega^1 \otimes \pi_1(a\varepsilon_1+b\varepsilon^2)
+ \Omega^2\otimes \pi_1(m\varepsilon_1+n\varepsilon^2)=
(a\Omega^1 +m\Omega^2)\otimes \varepsilon^1.
\]
Similarly,
\[
\pi_2\Omega=(b\Omega_1+n\Omega_2)\otimes \varepsilon^2.
\]
In particular, for our $\Omega$ we have simply
\[
\pi_1(F\otimes \varepsilon^1+*F\otimes\varepsilon^2)=
F\otimes\varepsilon^1,\
\pi_2(F\otimes\varepsilon^1+*F\otimes\varepsilon^2)
=*F\otimes\varepsilon^2.
\]
We consider now the expression $\vee(\delta\Omega,*\Omega)$. In our basis
$(\varepsilon^1,\varepsilon^2)$ we obtain
\[
\vee(\delta \Omega,*\Omega)=
\vee (\delta F \otimes\varepsilon^1 +\delta *F \otimes\varepsilon^2,
*F\otimes\varepsilon^1+**F\otimes\varepsilon^2)=
\]
\[
=(\delta F\wedge *F)\otimes\varepsilon^1\vee\varepsilon^1 +
(\delta *F\wedge **F)\otimes\varepsilon^2\vee\varepsilon^2+
(\delta **F\wedge F + \delta*F\wedge *F)
\otimes\varepsilon^1\vee\varepsilon^2.
\]
The first two components of this expression determine how much
energy-momentum the field is potentially able to exchange with the external
field, and the third component determines how much energy-momentum may be
redistributed between $F$ and $*F$. Hence, if the field $\Omega$ is free, all
these three components must be zero, and we obtain our free field equations
(recall $**F=-F$):
\begin{equation}
\delta F\wedge *F=0,\ \delta *F\wedge F=0,\         
\delta F\wedge F-\delta*F\wedge *F=0.
\end{equation}
In components, these equations in terms of the coderivative $\delta$ read
\begin{equation}
F_{\mu\nu}(\delta F)^\nu=0,\ (*F)_{\mu\nu}(\delta*F)^\nu=0,\
F_{\mu\nu}(\delta*F)^\nu+(*F)_{\mu\nu}(\delta F)^\nu=0.       
\end{equation}
These equations, in view of the energy-momentum relations
\begin{equation}
Q_\mu^\nu=\frac {1}{4\pi}\biggl[\frac{1}{4}
F_{\alpha\beta}F^{\alpha\beta}\delta_\mu^\nu-F_{\mu\sigma}
						F^{\nu\sigma}\biggr]=
\frac {1}{8\pi}\biggl[-F_{\mu\sigma}F^{\nu\sigma}-
(*F)_{\mu\sigma}(*F)^{\nu\sigma}\biggr],                  
\end{equation}
\noindent
and
\begin{equation}
\nabla_\nu Q_\mu^\nu=-
\frac {1}{4\pi}\biggl[F_{\mu\nu}(\nabla_\sigma F^{\sigma\nu})+
(*F)_{\mu\nu}(\nabla_\sigma (*F)^{\sigma\nu})\biggr]             
\end{equation}
in Maxwell theory make possible using the standard energy-momentum tensor of
Maxwell theory as energy-momentum tensor of this extended electromagnetic
theory.

In the same way, in terms of the exterior derivative $\mathbf{d}$ we have
\begin{equation}
(*F)^{\mu\nu}({\bf d}*F)_{\mu\nu\sigma}=0,
\ F^{\mu\nu}({\bf d}F)_{\mu\nu\sigma}=0,\
(*F)^{\mu\nu}({\bf d}F)_{\mu\nu\sigma}+                      
F^{\mu\nu}({\bf d}*F)_{\mu\nu\sigma}=0.
\end{equation}

The coordinate free form of (26) reads:
\begin{equation}
\vee(\delta\Omega,*\Omega)=0.                      
\end{equation}
The above equations are equivalent to the nonrelativistic equations
(15)-(17) when $F_{\mu\nu}$ are correspondingly interpreted.

Now, assume that $\Omega$ propagates in presence of an external field which
is able to exchange energy-momentum with $\Omega$. This means that the
external field must have its own "tools" to participate in the interaction. In
classical electrodynamics the usual external field is "continuously
distributed charged particles". Its interaction tools are the electric and
magnetic charges and currents, and the particles exchange energy-momentum
with the field along two independent ways: with $F$ by means of $j_e$ and
with $*F$ by means of $j_m$.  In the formal generalization (11)-(13) we
introduced 4 vector fields $\mathbf{a}^i$ and 4 functions $a^i$ to describe
the most general exchange process. So, we have to give the relativistic
equivalent of equations (11)-(13). This is achieved through introducing two
$\mathcal{R}^2$-valued differential 1-forms, denoted by
$\Phi=\alpha^1\otimes\varepsilon^1+\alpha^2\otimes\varepsilon^2$, and
$\Psi=\alpha^3\otimes\varepsilon^1+\alpha^4\otimes\varepsilon^2$, where
$\alpha^i, i=1,2,3,4$ are four 1-forms on $M$. The basic equation takes the
form
\begin{equation}
\vee(\delta \Omega,*\Omega)=
\vee(\Phi,*\pi_1\Omega)\ +\ \vee(\Psi,*\pi_2\Omega).       
\end{equation}
This equation is equivalent to (11)-(13) with
$\alpha^i=(\mathbf{a}^i,a^i)$: $\alpha^1$ describes the capability of the
external field to exchange energy-momentum with $F$, $\alpha^4$ describes the
the capability of the external field to exchange energy-momentum with $*F$,
and the couple ($\alpha^2, \alpha^3$) describes the capability of the
external field to affect the intrafield energy-momentum redistribution.  We'd
like to mention once again that (32), or (11)-(13), describe the most general
from formal point of view case, a concrete system may have, for example,
some of the 1-forms $\alpha^i$ equal to zero, or dependent on one another.

We give now the relativistic equivalent of equation (24). In components and
under the usual interpretation of $F_{\mu\nu}$ we have
\[
\Bigl[*(\delta F\wedge F)\Bigr]_\mu=
(\delta F)^\sigma(*F)_{\sigma\mu}=
\Biggl[\left({\rm rot}{\bf E}+\frac1c\frac{\partial
{\bf B}}{\partial t}\right)\times{\bf B}-
{\bf E}{\rm div}{\bf B},\
-\mathbf{E}.\left({\rm rot}\mathbf{E}+
\frac1c\frac{\partial \mathbf{B}}{\partial t}\right)\Biggr].
\]
Recalling relations (20)-(22) we see that the right-hand side of the above
relation is equal to
\[
\Bigl[\vec\mathcal{E}\times\mathbf{B}, -\vec\mathcal{E}.\mathbf{E}\Bigr]=
\Bigl[\vec\mathcal{E}\times\mathbf{B}, -\mathcal{H}\Bigr].
\]
Now, equation (24) is equivalent to
\begin{equation}
\mathbf{d}\left(\delta F\wedge F\right)=0.            
\end{equation}
If the components of $F$ are finite functions with respect to the spatial
coordinates $(x,y,z)$, which is possible in the nonlinear sector of solutions
of (32), then making use of Stokes' theorem, equation (33) leads to an
integral conserved quantity, which after some normalization appears as a
natural integral characteristic of the internal rotation (spin) properties of
the solution as it was explained in the nonrelativistic consideration.  We
note two things: first, no isometries are needed here to build this conserved
quantity since the energy-momentum tensor is not used; second, the local
expression of this conserved quantity depends on the derivatives of the field
functions while the energy-momentum densities do {\it not} depend on the
derivatives of $F_{\mu\nu}$.

\section{Conclusion}
The main purpose of this paper was to show that the duality properties of the
classical Maxwell equations together with the idea that the
local energy-momentumn balance relations should be the starting point for
basic theoretical assumptions, naturally lead to the nonlinear equations for
the electromagnetic field (in vacuum and in presence of external fields)
of Extended Electrodynamics. The full respect of duality we paid braught us
to a full formal equivalence of the two vector-components of the field
$\omega$, or $\Omega$, which was further recognised as a full
${\cal R}^2$-cavariance of the mathematical model object. We consider this
approach for nonlinearization of the field equations as an
appropriate extension of the second Newton's law of classical mechanics. Our
view on the free microobjects as obeying the Planck's relation $W=h\nu$
necessarily resulted in favoring the soliton concept as the most appropriate
working tool for now, because no point-like conception is able to explain the
availability of spin-momentum of photons. This reflects our view, based
on the conservation properties of the spin-momentum, that photons are real
objects and NOT theoretical imagination. So, photons must carry also
energy-momentum, their propagation in space must have some rotation-like
component, and at every moment they must occupy finite 3-volumes of definite
shape. Since we still don't know how this shape looks like we must keep
the possibility to make use of
{\it arbitrary} initial configurations, and our approach allows this through
the allowed arbitrariness of the amplitude function
$\Phi(x,y,\xi+\varepsilon z)$. This important moment allows the
well known localizing functions  from differential topology, used usually in
the partition of unity construction, to be used for making the spatial
dimensions of the solution FINITE, and NOT smoothly vanishing just at
infinity, as is the case of the usual soliton solutions.

Finally, we'd like to mention that we are allowed to choose the amplitude
function $\Phi$ of a "many-lump" kind, i.e. at every moment to be different
from zero inside many non-overlaping 3-regions, probably of the same shape,
so we are able to describe a flow of consistently propagating photons of the
same polarization and of the same phase. Some of such "many-lump" solutions
(flows) may give the impression of (parts of) plane waves.
\vskip 1cm
REFERENCES
\vskip 0.5cm
1. S.Donev, LANL e-print, hep-th/0006208.

2. S.Donev, M.Tashkova, Proc.Roy.Soc. of London A450, 281-291, 1995.

3. B.Lehnert, Physica Scripta, vol.T82, 89-94, 1999.

\end{document}